\documentstyle[prl,aps,preprint,tighten,floats,epsf,rotate]{revtex}

\newbox\rotbox

\begin{document}
\draft
\setcounter{page}{0}
\def\footnoterule{\kern-3pt \hrule width\hsize \kern3pt}
%
\title{New QCD Sum Rules for Nucleon Tensor Charge\thanks
{This work is supported in part by funds provided by the U.S.
Department of Energy (D.O.E.) under cooperative 
research agreement \#DF-FC02-94ER40818.}}

\author{Xuemin Jin\footnote%
{jin@ctpa02.mit.edu}
 and Jian Tang\footnote%
{jtang@ctpa02.mit.edu}
}

\address{Center for Theoretical Physics \\
Laboratory for Nuclear Science \\
and Department of Physics \\
Massachusetts Institute of Technology\\
Cambridge, Massachusetts 02139, USA \\}

\date{MIT-CTP-2634 ~~~~~ hep-th/9705269 {~~~~~} May 1997}
\maketitle

\thispagestyle{empty}

\begin{abstract}

Two new QCD sum rules for nucleon tensor charge are derived from a
mixed correlator of spin-1/2 and spin-3/2 nucleon interpolating fields.
These sum rules are analyzed along with a sum rule obtained from 
the usual correlator of a general spin-1/2 nucleon interpolating field.
The validity and reliability of the sum rules are examined by 
monitoring the contaminations arising from the transitions and continuum
and the convergence of the operator product expansion. Valid sum rules
are identified and their predictions are presented. It is found that
the vacuum tensor susceptibility induced by the external field plays 
an important role in determining both the validity and predictions 
of the sum rules. The uncertainties associated with the sum-rule 
predictions are also discussed.

\end{abstract}

\vspace*{\fill}
\begin{center}
Submitted to: {\it Physical Review D}
\end{center}


\newpage
\section{Introduction}
\label{intro}

Tensor charge of the nucleon, defined through the nucleon matrix
element of the tensor current, $\overline{q} \sigma_{\alpha\beta} q$,
at zero momentum transfer, is perhaps the least known nucleon charge both 
theoretically and experimentally. Like other charges of the nucleon, 
the tensor charge reveals important information on the nonperturbative 
structure of the nucleon and understanding it from the underlying theory 
of strong interaction, quantum chromodynamics (QCD), is of great importance.

The nucleon tensor charge $\delta q$ is defined by
\begin{equation}
\langle N|\,\overline{q}\, \sigma_{\alpha\beta}\, q\,|N\rangle
= \delta q\, \overline{U}\,\sigma_{\alpha\beta}\,U\ ,
\label{deltaq-def}
\end{equation}
where $U$ stands for the nucleon Dirac spinor. 
It has been shown by Jaffe and Ji \cite{jaffe91} that the first moment 
of the twist-two transversity quark distribution in a nucleon, $h_1(x)$, 
is related to the nucleon tensor charge
\begin{equation}
\int_0^1 dx \left[
h^q_1(x) - \overline{h}^q_1(x)\right] = \delta q\ ,
\label{chr-h}
\end{equation}
where $h_1^q$ and $\overline{h}_1^q$ receive contributions from quark
and antiquark (of flavor $q$), respectively.  
Unlike the twist-two spin averaged quark distribution $f_1(x)$ and  
helicity difference quark distribution $g_1(x)$, the $h_1(x)$ flips 
chirality and is thus suppressed in inclusive deep inelastic 
lepton scattering \cite{jaffe96}. As such, there is no existing 
experimental data on $h_1(x)$ and the nucleon tensor charge. 
However, there is no such suppression of $h_1(x)$ in deep inelastic 
processes with hadronic initial states such as Drell-Yan 
\cite{ralston79,jaffe96}, and  experiments have been planned 
to measure the quark transversity distribution in the nucleon at 
RHIC \cite{rhic}, HERA \cite{hera}, and CERN \cite{cern}.

Recently, the nucleon tensor charge has been investigated theoretically 
in various models and approaches \cite{he95,he96,kim96,schmidt97,aoki96}.
Previous studies of the nucleon tensor charge via the QCD sum-rule 
method have been made by He and Ji \cite{he95,he96}. These 
authors have found that the sum rules obtained from the correlator of Ioffe's
nucleon interpolating field \cite{ioffe81} are unstable, making the 
extraction of the nucleon tensor charge from those sum rules  
difficult. In this paper, we derive two new QCD sum rules from the 
mixed correlator of spin-1/2 and spin-3/2 interpolating fields 
using the external-field method. The consideration of the mixed
correlator is motivated by its success in nucleon mass sum 
rules \cite{leinweber97} and in the sum rules for nucleon 
axial-vector coupling constants \cite{lee97}. We analyze 
these new sum rules along with a sum rule obtained from 
the usual correlator with a  general spin-1/2 nucleon 
interpolating field. We examine the performances and reveal 
the validity and reliability of these sum rules.

A well-known problem that arises in calculating hadronic matrix 
elements within the QCD sum-rule method is the unwanted physics 
associated with transitions from ground state to excited 
states. The contributions of these transitions are not exponentially
suppressed relative to the ground state contribution which contains 
the ground state property of interest \cite{ioffe84}.  In practice, 
there are two formalisms for treating the transitions. The usual formalism 
is to model the sum over all the transitions  from ground state 
to excited states by a single unknown 
parameter. This parameter is extracted from the sum rules, 
along with the ground state property, and the continuum threshold 
is usually fixed at the same value as that for the corresponding 
mass sum rule.  The resolution of the ground state signal relies 
on the different polynomial (in Borel mass) behaviors of the ground 
state signal and the transition term.

An alternative formalism for treating the transitions is to multiply the
invariant functions by the factor $M_h^2 - q^2$ (with $M_h$ the ground
state hadron mass) before performing the Borel transform. The transition
contribution is then exponentially suppressed relative to the ground
state contribution and can be included in the continuum model in practice.
In this case, it is important to treat the continuum threshold as an
unknown parameter to be extracted from the sum rule
along with the ground state property. Assuming the threshold to be the
same as that for the mass sum rule is unjustified and may introduce
artificial bias to the extracted hadron properties \cite{jin97}. 
We observe that in the previous works \cite{he95,he96} 
the alternative formalism was adopted but the continuum threshold 
was taken to be the same as the one used in the nucleon mass sum rule. 
In the present paper, we shall consider both formalisms.

Both sides of QCD sum rules depend on an auxiliary parameter--Borel mass,
which is introduced through the use of the Borel transform. If a sum rule 
were perfect, one would expect that the two sides of the sum rule overlap for
all values of the Borel mass. However, in practice,
the two sides of the sum rules overlap only in a limited
range of the Borel mass (at best) because of the truncation
of the operator product expansion (OPE) and the crudity of the 
models for the transitions and continuum. 
Thus, in order to extract the properties of the ground state  
by matching the sum rules, one should work in a 
region of Borel mass where the ground state contribution  
dominates the phenomenological side. This usually sets an upper bound 
in the Borel mass space, beyond which the excited-state and transition
contributions dominate and the background noise is hence too strong to 
reliably isolate the ground state signal. 
On the other hand, the truncated OPE must be sufficiently convergent 
as to accurately describe the true OPE. This, in practice, sets a lower limit
in the Borel mass space, beyond which higher order terms not present
in the truncated OPE may be out of control. 

Therefore, we expect a sum rule to work if the two sides of the sum rule match
in a {\it valid} window in Borel mass space where the ground state 
contribution of interest dominates the phenomenological side and the 
higher order OPE terms are under control. The validity and 
reliability of a sum rule is then determined by the quality of overlap 
of the two sides, the size of valid Borel window, and the relative 
contributions of the excited states and the highest OPE terms 
in the valid Borel window. A sum rule failing to have a valid Borel window
should be considered invalid, and results obtained from such an invalid 
sum rule may be meaningless and misleading.  In this paper, we will 
analyze the sum rules with respect to the above criteria.

This paper is organized as follows. In Sec.~\ref{sumrule}, we derive 
QCD sum rules for the nucleon tensor charge within the external-field 
approach.  In Sec.~\ref{analysis}, we analyze the sum rules and present 
results and discussions. Section~\ref{summary} is devoted to summary 
and conclusion.
\section{Sum rules for nucleon tensor charge}
\label{sumrule}

In this section, we first give a brief outline of the external-field 
QCD sum-rule method with an emphasis on some issues related to the
treatment of the transitions between ground state and excited
states. We then use this method to derive QCD sum rules for the nucleon 
tensor charge.

\subsection{Outline of the method}

Consider the nucleon two-point correlator in the presence 
of a constant external (antisymmetric) tensor field, 
$Z^{\alpha\beta}$, defined by
\begin{equation}
\Pi(q) = i\int d^4x e^{iq\cdot x}
\langle 0|T \eta(x)\overline{\eta}(0)|0\rangle_{_Z}\ ,
\label{corr-gen}
\end{equation}
where $\eta$ is a nucleon interpolating field, constructed from quark
fields, which couples to the nucleon. 
To lowest dimension, there are two independent interpolating fields
which couple to spin-1/2 states only and carry the quantum numbers of
the nucleon. So, a general spin-1/2 interpolating field for the proton can be
expressed as
\begin{equation}
\eta^{1/2} = \eta_1 + \beta \eta_2\ ,
\label{eta-1/2}
\end{equation}
where $\beta$ is a real parameter and
\begin{eqnarray}
\eta_1(x) &=& \epsilon_{abc}\left[
u^{aT}(x) C\gamma^5 d^b(x)\right] u^c(x)\ ,
\\*[7.2pt]
\eta_2(x) &=& \epsilon_{abc}\left[
u^{aT}(x) C d^b(x)\right]
\gamma_5 u^c(x)\ .
\end{eqnarray}
Here $u(x)$ and $d(x)$ denote up and down quark operators,
$a, b,$ and $c$ are color indices, and $C$ is the charge conjugation
matrix. The interpolating field for the neutron can be obtained by
changing $u(d)$ into $d(u)$. The interpolating field advocated by 
Ioffe \cite{ioffe81}  and used in previous sum-rule calculations 
of the nucleon tensor charge \cite{he95,he96} may be recovered by
setting $\beta = -1$ and multiplying an overall factor of $-2$.

There is a spin-3/2 interpolating field which also couples to the proton 
through its spin-1/2 component \cite{chung82,leinweber97}
\begin{equation}
\eta_\mu^{3/2} = \epsilon_{abc}
\left[\left(u^{aT} C \sigma_{\rho\lambda} d^b\right)
\sigma^{\rho\lambda}\gamma_\mu u^c -
\left(u^{aT} C \sigma_{\rho\lambda} u^b\right)
\sigma^{\rho\lambda} \gamma_\mu d^c\right]\ .
\label{eta-3/2}
\end{equation}
The couplings of $\eta^{1/2}$ and $\eta^{3/2}_\mu$ to the proton
are defined as
\begin{equation}
\langle 0|\eta^{1/2}|N\rangle = \lambda_1 U(q) \ ,\hspace*{1cm}
\langle N|\overline{\eta}^{3/2}_\mu |0\rangle
= \lambda_2 \overline{U}(q)\gamma_5 \left({4 q_\mu\over M_N }+ \gamma_\mu\right)\ ,
\label{couplings}
\end{equation}
where $U(q)$ denotes the Dirac spinor of the proton with the normalization
$\overline{U}(q)U(q) = 2 M_N$, and $\lambda_1$ and $\lambda_2$ describe
the coupling strengths of the interpolating fields to the proton.

Therefore, independent QCD sum rules can be obtained from the correlators
of various combinations of $\eta^{1/2}$ and  $\eta_\mu^{3/2}$.
In the present work, we will consider the mixed correlator of
$\eta^{1/2}$ and  $\eta_\mu^{3/2}$ as well as the 1/2 correlator
of two $\eta^{1/2}$'s. 
While the latter, at $\beta = -1$, leads to the sum rules discussed
previously, the former gives rise to new independent sum rules
for the nucleon tensor charge. In Ref~\cite{leinweber97}, it has been 
demonstrated that the nucleon mass sum rules from the mixed correlator 
have broader valid Borel windows and relatively smaller continuum 
contamination as compared to the sum rules from the 1/2 correlator.

The subscript $Z$ in Eq.~(\ref{corr-gen})
denotes that $\Pi(q)$ is evaluated by adding a tensor coupling term
\begin{equation}
\Delta{\cal L} = Z^{\alpha\beta} \sum_{q} \, g_q\, \overline{q}(x) 
                 \,\sigma_{\alpha\beta} \, q(x)\ ,
\label{dl}
\end{equation}
to the QCD Lagrangian, where $q= \{u,d,s\}$. Here $g_q$ keeps track of
the flavor structure of the external field. For example, if one is 
interested in the isovector (isoscalar) tensor charge, 
$g^v_{_{\rm T}} \equiv \delta u -\delta d$ ($g^s_{_{\rm T}}
\equiv \delta u + \delta d$), one simply sets $g_u=-g_d=1$
($g_u = g_d =1$) and $g_s=0$. The external-field method proceeds 
by expanding $\Pi(q)$ to first order in the external field
\begin{equation}
\Pi(q) = \Pi^{(0)}(q) + Z^{\alpha\beta}\, \Pi^{(1)}_{\alpha\beta}(q) 
+ \cdot\cdot\cdot\ ,
\label{corr-gen-exp}
\end{equation}
where $\Pi^{(0)}(q)$ is the correlator in the absence of 
the external field from which the usual nucleon mass sum rules are derived. 
It is the linear response to the external field, $\Pi^{(1)}_{\alpha\beta}(q)$, 
that gives rise to the sum rules for the nucleon tensor charge.

The QCD side of the sum rule is obtained by carrying out the OPE. 
The external field can interact
with the quark field directly; it also polarizes the QCD vacuum and induces
nonperturbative condensates. The latter are summarized by vacuum 
susceptibilities, which, to first order in the external field, are 
defined as \cite{he95,he96}
\begin{eqnarray}
& &\langle\, \overline{q}\,\sigma_{\alpha\beta}\, q\,\rangle_{_Z}
   \,\equiv g_q\, \chi\, \langle\, \overline{q} q\, \rangle Z_{\alpha\beta}\ ,
\label{sus-chi}
\\*[7.2pt]
& &\langle\, \overline{q}\, g_s G_{\alpha\beta} \,q\,\rangle_{_Z}
  \,\equiv g_q\, \kappa\, \langle\, \overline{q} q\,\rangle Z_{\alpha\beta}\ ,
\label{sus-kappa}
\\*[7.2pt]
& &\langle\, \overline{q}\, g_s \widetilde{G}_{\alpha\beta} \gamma_5\, q\,
\rangle_{_Z}\, \equiv -i g_q\, \xi\, \langle\, \overline{q} q\,
\rangle Z_{\alpha\beta}\ ,
\label{sus-xi}
\end{eqnarray}
where $g_s$ is the strong coupling constant ($D_\mu = \partial_\mu
+ i g_s {\cal A}_\mu$), $\widetilde{G}_{\alpha\beta}\equiv {1\over 2}
\epsilon_{\alpha\beta\rho\lambda} G^{\rho\lambda}$ with $G_{\alpha\beta}$
the gluon field tensor, $\chi$, $\kappa$, and $\xi$ are vacuum 
susceptibilities, and $\langle \hat{O}\rangle$ denotes the usual vacuum
condensate. Here, the isospin
breaking effect has been neglected and will be henceforth. The strangeness
contribution will also be ignored in the present paper as it is expected
to be small \cite{kim96,aoki96}. The calculation of the Wilson
coefficients is straightforward following the techniques discussed 
extensively in the literature. Some details of the calculations are 
presented in the Appendix.

We now turn to the phenomenological description, which can be obtained by 
expanding $\Pi^{(1)}_{\alpha\beta}(q)$ in terms of physical intermediate 
states. For a generic invariant function, the phenomenological representation 
can be written as 
\begin{eqnarray}
\Pi^{(1)}(q^2) &\sim& {\lambda_{\cal O}^2\,  g_{_{\rm T}} 
\over (M_N^2 - q^2)^2}
+ \sum_{N^*}{C_{N\leftrightarrow N^*}
\over (M_N^2 - q^2) (M_{N^*}^2 - q^2)}
\nonumber
\\*[7.2pt]
& &\hspace*{1cm}
+ \mbox{terms involving only excited states}\ ,
\label{old-exp}
\end{eqnarray}
where $M_N$ and $M_{N^*}$ are the masses of the ground state nucleon and 
the excited state, respectively, and $\lambda_{\cal O}^2  =
\lambda_1\lambda_2,\, \lambda_1^2$ for the mixed and 1/2 correlators,
respectively. The first term contains the 
nucleon tensor charge $g_{_{\rm T}} = \sum_{q} g_q\,\delta q$ of interest, 
the second term is the transition term from ground state to excited states, 
and the rest involves only the excited states. After a usual Borel 
transformation, one finds
\begin{eqnarray}
\Pi^{(1)}(M^2) &\sim&
{\lambda_{\cal O}^2\over M^2} g_{_{\rm T}}\, e^{-M_N^2/M^2}
+  \sum_{N^*} \Biggl[
{C_{N\leftrightarrow N^*}\over M_{N^*}^2-M_N^2}\,
\left(1-e^{-(M_{N^*}^2 - M_N^2)/M^2}\right)\Biggr]\, e^{-M_N^2/M^2}
\nonumber
\\*[7.2pt]
& &\hspace*{1cm}
+  \mbox{exponentially suppressed terms}\ ,
\label{old-form}
\end{eqnarray}
where $M$ is the Borel mass.
It can be seen clearly that the second term is not 
exponentially damped as compared to the first term. The usual formalism is 
to approximate the second term as $A e^{-M_N^2/M^2}$ with $A$ 
a phenomenological constant to be extracted from the sum rule
along with $g_{_{\rm T}}$.

An alternative formalism is to  use the combination 
$(M_N^2 - q^2) \,\, \Pi^{(1)}(q^2)$, instead of $\Pi^{(1)}(q^2)$ alone. 
The phenomenological side then becomes
\begin{eqnarray}
\widetilde{\Pi}^{(1)}(M^2) \sim 
\lambda_{\cal O}^2 g_{_{\rm T}} e^{-M_N^2/M^2}
+ \sum_{N^*} C_{N\leftrightarrow N^*}
e^{-M_{N^*}^2/M^2}
+\mbox{exponentially suppressed terms}\ .
\end{eqnarray}
The transitions from ground state to excited states
(second term) are now exponentially suppressed with respect to the 
ground state (first) term and hence can be absorbed into the continuum
model. As such, one no longer needs to introduce any phenomenological 
parameter to represent the transitions from the ground state to excited 
states as their contribution has been included in the continuum model. 
However, as stressed in Ref.~\cite{jin97} the continuum threshold must 
be treated as an unknown parameter to be extracted from the sum rule.

Since the linear response $\Pi^{(1)}_{\alpha\beta}$ in general contains 
distinct Dirac and Lorentz structures, one may obtain many 
sum rules, one for each invariant structure. However, these sum rules
do not work equally well in practice. In particular, some sum rules
work well while the others may fail. Recently, it has been pointed
out by the present authors \cite{jin97a} that chirality plays an important 
role in determining the reliability of baryon sum rules. There it was
argued that for light baryons the chiral-odd sum rules (where the 
chiral-odd operators dominate) are generally more reliable than the 
chiral-even sum rules (where the chiral-even operators dominates). 
Thus, we will only consider the chiral-odd sum rules and disregard
the chiral-even sum rules in the discussions to follow.

\subsection{Sum rules from the mixed correlator}

The mixed correlator is defined by
\begin{eqnarray}
\Pi_{\mu\nu}(q) &=& i \int d^4x\, e^{i q\cdot x}
\langle 0| T[ \eta^{1/2}_\mu (x) 
\overline{\eta}^{3/2}_\nu(0) ]
|0\rangle_{_Z}
\nonumber
\\
&=& \Pi^{(0)}_{\mu\nu}(q)+ Z^{\alpha\beta} \Pi^{(1)}_{\mu\nu,\alpha\beta}(q)
+\cdots\ ,
\label{corr-mix}
\end{eqnarray}
where $\eta^{1/2}_\mu\, \equiv \gamma_\mu\gamma_5 \eta^{1/2}$
and $\eta^{1/2}$ and $\eta^{3/2}_\nu$ are given in Eqs.~(\ref{eta-1/2})
and (\ref{eta-3/2}), respectively. The use of $\eta^{1/2}_\mu$ is
to facilitate the calculations and does not affect the sum rules.
The linear response of $\Pi_{\mu\nu}(q)$ to the external field
can be decomposed into various invariant structures:
\begin{eqnarray}
\Pi^{(1)}_{\mu\nu,\alpha\beta}(q)
= & &\Pi_1(q^2)\, \gamma_\mu \left(
\hat{q}\sigma_{\alpha\beta}+\sigma_{\alpha\beta}\hat{q}
\right) \gamma_\nu
+\Pi_2(q^2)\, \gamma_\mu \sigma_{\alpha\beta} q_\nu
\nonumber
\\
& &
+\Pi_3(q^2)\, i \gamma_\mu \left(
q_\alpha\gamma_\beta - q_\beta\gamma_\alpha
\right) \hat{q} q_\nu
+\cdots\ ,
\label{mix-dec}
\end{eqnarray}
where $\hat{q}\equiv q_\mu\gamma^\mu$. Here, the first three structures
receive contributions from the ground state proton and lead to 
chiral-odd sum rules. Since there is no continuum model for 
the sum rule from $\Pi_3$, we only focus on the sum rules from 
$\Pi_1$ and $\Pi_2$. Adopting the usual formalism for the transitions,
we obtain the sum rule from $\Pi_1$
\begin{eqnarray}
& &{a \over 8}\, c_1 E_0 L^{4/9} M^2
-{m_0^2 a\over 24}\, c_2 L^{-2/27}
    \left[\ln\left({M^2\over\Lambda^2}\right) 
             - \gamma_{_{\rm EM}}\right]
+{m_0^2 a \over 96}\, c_3 L^{-2/27}
\nonumber
\\*[7.2pt]
& &-{\chi a^2 \over 24}\, c_4 L^{8/27}
-{\kappa a^2 \over 144 M^2}\, c_5 L^{8/9}
-{\xi a^2 \over 72 M^2} \, c_6 L^{8/9}
+{\chi m_0^2 a^2 \over 1728 M^2}\, c_7 L^{-6/27}
\nonumber
\\*[7.2pt]
& &+{a b \over 864 M^2}\, c_8 L^{4/9}
= M_N \widetilde{\lambda}_1\widetilde{\lambda}_2
\left({g_{_{\rm T}}\over M^2} + A_1\right) e^{-M_N^2/M^2} \ ,
\label{sr-mix-1}
\end{eqnarray}
and the sum rule from $\Pi_2$
\begin{eqnarray}
& &{2 a\over 3}\, d_1 E_0 L^{4/9} M^2
-{m_0^2 a \over 3}\, d_2 L^{-2/27}
  \left[\ln\left({M^2\over\Lambda^2}\right)
   -\gamma_{_{\rm EM}}\right] 
- {m_0^2 a\over 6}\, d_3 L^{-2/27}
\nonumber
\\*[7.2pt]
& &
+{\chi a^2 \over 3}\, d_4 L^{8/27}
+{\kappa a^2 \over 9 M^2}\, d_5 L^{8/9} 
+{\xi a^2 \over 9 M^2}\, d_6  L^{8/9} 
-{\chi m_0^2 a^2 \over 216  M^2}\, d_7 L^{-6/27}
\nonumber
\\*[7.2pt]
& & +{a b \over 216  M^2}\, d_8 L^{4/9} 
= 4 {\widetilde{\lambda}_1\widetilde{\lambda}_2 \over M_N}
\left[ g_{_{\rm T}}\left({2 M_N^2\over M^2} -1\right)+A_2\right] e^{-M_N^2/M^2}\ ,
\label{sr-mix-2}
\end{eqnarray}
where $a = -(2\pi)^2 \langle\overline{q} q\rangle$,
$b = \langle g_s^2 G^2\rangle$,
$m_0^2 = -\langle\overline{q} g_s\sigma\cdot G q\rangle
/\langle\overline{q} q\rangle$, 
$\widetilde{\lambda}_1 = (2\pi)^2\lambda_1$, 
$\widetilde{\lambda}_2 = (2\pi)^2 \lambda_2$, 
$E_0 = 1-e^{-s_0/M^2}$ with $s_0$ the continuum threshold, and 
$\gamma_{_{\rm EM}} = 0.577\cdots$ the Euler-Mascheroni constant. 
The anomalous dimensions of various operators have been 
taken into account through the factor 
$L = \ln(M^2/\Lambda_{\rm QCD}^2)/\ln(\mu^2/\Lambda_{\rm QCD}^2)$,
where $\mu = 500 {\rm MeV}$ is the renormalization scale and 
$\Lambda_{\rm QCD} = 150 {\rm MeV}$ is the QCD scale parameter.
Here $\Lambda$ denotes the infrared cut-off, which arises from the
quark propagator in the presence of both the external field and 
background gluon field (see Appendix); a reasonable choice for 
$\Lambda$ is $\Lambda = \mu$ \cite{wilson87}. The $c_i$ and $d_i$ 
are defined as  
\begin{eqnarray}
& & c_1 = (5+9\beta) g_u+(1-9\beta) g_d\ ,\hspace*{2cm}
    c_2 = (1-\beta) (4 g_u-g_d)\ ,
\nonumber
\\*[7.2pt]
& & c_3 = -(29+85\beta) g_u+(5-71\beta) g_d\ ,\hspace*{1.1cm}
    c_4 = -7(1+\beta) g_u+(1-11\beta) g_d\ ,
\nonumber
\\*[7.2pt]
& & c_5 = 69 \beta g_u+ (-6+2\beta) g_d\ , \hspace*{2.6cm}
    c_6 = (7+9\beta) g_u - (1+21\beta) g_d\ ,
\nonumber
\\*[7.2pt]
& & c_7 = -(47+87\beta) g_u +(23-237\beta) g_d\ ,\hspace*{0.7cm}
    c_8 = 5 (1+\beta) g_u- (8-28\beta) g_d\ ,
\nonumber
\\*[7.2pt]
& & d_1 = (3+4\beta)g_u+5\beta g_d\ ,\hspace*{3.1cm}
    d_2 =(1-\beta) (g_u-g_d)\ ,
\nonumber
\\*[7.2pt]
 & & d_3 = (2+13\beta) g_u+(1+11\beta) g_d\ ,\hspace*{1.7cm}
     d_4 = (7+3\beta) g_u-(1-3\beta) g_d\ ,
\nonumber
\\*[7.2pt]
 & &  d_5 = (10-13\beta) g_u - (4+2\beta) g_d\ ,\hspace*{1.7cm}
      d_6 = 2\beta g_u+6\beta g_d\ ,
\nonumber
\\*[7.2pt]
 & & d_7 = -(158+10\beta) g_u +(7+71\beta) g_d\ ,\hspace*{0.9cm}
     d_8 = (5+7\beta) g_u +(1+23\beta) g_d\ .
\label{cds}
\end{eqnarray}
When the alternative formalism for the transitions is used, the
sum rule from $\Pi_1$ is
\begin{eqnarray}
& &{a \over 8}\, c_1 \left(M_N^2 M^2 E_0-M^4 E_1\right) L^{4/9}
-{m_0^2 a \over 24}\, c_2 \left[M_N^2\,\ln\left({M^2\over\Lambda^2}\right)
-M_N^2\, \gamma_{_{\rm EM}}-M^2 E_0\right] L^{-2/27}
\nonumber
\\*[7.2pt]
 & &
+{m_0^2 a\over 96}\, c_3 M_N^2 L^{-2/27}
-{\chi a^2 \over 24}\, c_4 M_N^2 L^{8/27}
-{\kappa a^2 \over 144}\, c_5 \left(1+{M_N^2\over M^2}\right) L^{8/9}
\nonumber
\\*[7.2pt]
 & &
-{\xi a^2 \over 72}\, c_6 \left(1+{M_N^2\over M^2}\right) L^{8/9}
+{\chi m_0^2 a^2 \over 1728}\, c_7 \left(1+{M_N^2\over M^2}\right) L^{-6/27}
\nonumber
\\*[7.2pt]
 & &
+{a b \over 864}\, c_8 \left(1+{M_N^2\over M^2}\right) L^{4/9}
= M_N \widetilde{\lambda}_1\widetilde{\lambda}_2\, 
g_{_{\rm T}} e^{-M_N^2/M^2} \ ,
\label{sr-mix-1a}
\end{eqnarray}
and the sum rule from $\Pi_2$ is
\begin{eqnarray}
& &{2 a \over 3}\, d_1 \left(M_N^2 M^2 E_0-M^4 E_1\right) L^{4/9}
-{m_0^2 a \over 3}\, d_2 \left[M_N^2\, \ln\left({M^2\over\Lambda^2}\right)
-M_N^2\, \gamma_{_{\rm EM}}-M^2 E_0\right] L^{-2/27}
\nonumber
\\*[7.2pt]
 & &
-{m_0^2 a \over 6}\, d_3 M_N^2 L^{-2/27}
+{\chi a^2 \over 3}\, d_4 M_N^2 L^{8/27}
+{\kappa a^2 \over 9}\, d_5 \left(1+{M_N^2\over M^2}\right) L^{8/9}
\nonumber
\\*[7.2pt]
 & &
+{\xi a^2 \over 9}\, d_6 \left(1+{M_N^2\over M^2}\right) L^{8/9}
-{\chi m_0^2 a^2 \over 216}\, d_7 \left(1+{M_N^2\over M^2}\right) L^{-6/27}
\nonumber
\\*[7.2pt]
 & &
+{a b \over 216}\, d_8 \left(1+{M_N^2\over M^2}\right) L^{4/9}
= 8 M_N \widetilde{\lambda}_1\widetilde{\lambda}_2\, g_{_{\rm T}} e^{-M_N^2/M^2} \ ,
\label{sr-mix-2a}
\end{eqnarray}
where $E_1 = 1-e^{-s_0/M^2} (s_0/M^2+1)$.

\subsection{Sum rules from the 1/2 correlator}

The 1/2 correlator is given by
\begin{eqnarray}
\Pi(q) &=& i \int d^4x\, e^{i q\cdot x}
\langle 0| T[ \eta^{1/2} (x) 
\overline{\eta}^{1/2}(0) ]
|0\rangle_{_Z}
\nonumber
\\
&=&\Pi^{(0)}(q) + Z^{\alpha\beta} \Pi^{(1)}_{\alpha\beta}(q)
+ \cdots\ .
\label{corr-usual}
\end{eqnarray}
The linear response term  in this case has three
distinct structures
\begin{equation}
\Pi^{(1)}_{\alpha\beta} = 
\Pi_4(q^2)\, \left(\hat{q}\sigma_{\alpha\beta} 
+ \sigma_{\alpha\beta}\hat{q}\right)
+\Pi_5(q^2)\,  \sigma_{\alpha\beta}
+\Pi_6(q^2)\,  
i\left(q_\alpha\gamma_\beta-q_\beta\gamma_\alpha\right)\hat{q}\ .
\label{usual-dec}
\end{equation}
However, only $\Pi_4$ leads to a chiral-odd sum rule, which is given by
\begin{eqnarray}
& & {a \over 32}\, f_1  L^{4/27} E_0 M^2
+{m_0^2 a \over 96}\, f_2 \left[\ln\left({M^2\over\Lambda^2}\right)
-\gamma_{_{\rm EM}}\right] L^{-10/27}
-{m_0^2 a \over 192}\, f_3 L^{-10/27}
\nonumber
\\*[7.2pt]
 & &-{\chi a^2 \over 48}\, f_4 
+{\kappa a^2 \over 576 M^2}\, f_5 L^{16/27}
+{\xi a^2 \over 96  M^2}\, f_6 L^{16/27} 
-{\chi m_0^2 a^2 \over 1728 M^2}\, f_7 L^{-14/27}
\nonumber
\\*[7.2pt]
 & &
+{a b \over 6912 M^2}\, f_8  L^{4/27} 
= \widetilde{\lambda}_1^2 \left({M_N\over M^2} g_{_{\rm T}} + A_4\right)
e^{-M_N^2/M^2}\ ,
\label{sr-usual}
\end{eqnarray}
in the usual formalism and
\begin{eqnarray}
& &{a \over32}\, f_1 L^{4/27} (M_N^2 E_0 M^2 - E_1 M^4)
+{m_0^2 a \over 96}\, f_2 
\left[M_N^2\, \ln\left({M^2\over\Lambda^2}\right)- 
M_N^2\, \gamma_{_{\rm EM}}
-E_0 M^2\right] L^{-10/27}
\nonumber
\\*[7.2pt]
 & &-{m_0^2 a \over 192}\, f_3 M_N^2 L^{-10/27}
-{\chi a^2 \over 48}\, f_4 M_N^2  
+{\kappa a^2 \over 576}\, f_5 L^{16/27} \left(1+{M_N^2\over M^2}\right)
\nonumber
\\*[7.2pt]
 & &+{\xi a^2 \over 48}\, f_6 L^{16/27} \left(1+{M_N^2\over M^2}\right)
-{\chi m_0^2 a^2 \over 1728}\, f_7 L^{-14/27} 
\left(1+{M_N^2\over M^2}\right)
\nonumber
\\*[7.2pt]
& &
+{a b\over 6912}\, f_8  L^{4/27} \left(1+{M_N^2\over M^2}\right)
= M_N \widetilde{\lambda}^2_1\, g_{_{\rm T}} e^{-M_N^2/M^2}\ ,
\label{sr-usual-a}
\end{eqnarray}
in the alternative formalism, where
\begin{eqnarray}
& & f_1 = (10+6\beta^2) g_u - (3+2\beta-\beta^2) g_d\ ,\hspace*{2cm}
    f_2 = (1-\beta)^2 g_u\ ,
\nonumber
\\*[7.2pt]
 & & f_3 = (4-10\beta+6\beta^2) g_u - 5(1-\beta^2) g_d\ ,\hspace*{1.8cm}
     f_4 = -2(1-\beta) g_u + (1-\beta^2) g_d\ ,
\nonumber
\\*[7.2pt]
 & & f_5 = (20+2\beta-22\beta^2) g_u +9(1-\beta^2) g_d\ ,\hspace*{1.6cm}
     f_6 = 2 (1-\beta) g_u - (1-\beta^2) g_d\ ,
\nonumber
\\*[7.2pt]
 & & f_7 = (6-7\beta+\beta^2) g_u - 16 (1-\beta^2) g_d\ ,
\nonumber
\\*[7.2pt]
 & &  f_8 = (62+36\beta+70\beta^2) g_u +(1-6\beta+17\beta^2) g_d\ .
\label{fs}
\end{eqnarray}

The sum rule Eq.~(\ref{sr-usual-a}), at $\beta = -1$, can be compared
with Eq.~(20) of  Ref.~\cite{he96}. While the first and fourth 
terms agree, the rest disagree. The factor $M_N^2 \ln(M^2/\Lambda^2)
-M_N^2 \gamma_{_{\rm EM}}- M^2 E_0$ in the second term has the form 
$M_N^2 \ln(M_N^2/\Lambda^2) -M_N^2$ in Ref.~\cite{he96}. We have also 
compared the corresponding form of this term before the Borel transformation
and found an overall sign difference. Furthermore,  
the Borel transformation of this term has not been done correctly 
in Ref.~\cite{he96} (see Appendix for more details). 
The third term is absent in Ref.~\cite{he96}.
 The differences in the last four terms arise from the 
incomplete calculation of the Wilson coefficients in Ref.~\cite{he96}
(see Appendix). There is also a minor difference in the 
anomalous dimension for the mixed quark-gluon operator 
$\overline{q} g_s \sigma \cdot G q$. We use $-2/27$, 
while Ref.~\cite{he96} simply takes 0.  
\section{Sum-rule analysis}
\label{analysis}

We now analyze the sum rules derived in the previous 
section. To this end, we optimize the fit of the two sides 
of each individual sum rule in a valid Borel window, which 
is chosen such that the highest order (in $1/M^2$) term 
contributes no more than $\sim 10\%$ to the QCD side while 
the  contribution of the transition plus continuum  
is less than 50\% of the total phenomenological side (i.e., the sum 
of the contributions from ground state, transition, and 
continuum). Those sum rules which fail to have a valid Borel 
window is considered invalid and useless. Such criteria have been
adopted in various sum-rule calculations \cite{leinweber97,%
lee97,jin95}. While the selection
of $50\%$ is obvious for the ground state dominance, the selection
of $10\%$ is a reasonably conservative criterion.
To quantify the fit of the two sides, we sample the sum rule
in the valid Borel window and minimize the measure $\delta (M^2)
=(\mbox{LHS-RHS})^2$ averaged over 50 points evenly spaced in the 
valid Borel window. Here, LHS and RHS denote the left- and right-hand 
sides of the sum rule, respectively.
\subsection{Inputs}

The inputs for the sum rules can be classified into three
categories: a) the usual vacuum condensates; b) the vacuum 
susceptibilities $\chi$, $\kappa$, and $\xi$; c) the nucleon mass $M_N$, 
the coupling strengths $\lambda_1$ and $\lambda_2$, and the continuum 
threshold $s_0$. [The continuum threshold is needed as input for the sum 
rules (\ref{sr-mix-1}--\ref{sr-mix-2}) and (\ref{sr-usual}).]
For the values of vacuum condensates we use the central values given
in Ref.~\cite{leinweber97}: $a = 0.52$ GeV$^3$,
$b = 1.2$ GeV$^4$, and $m_0^2 = 0.72$ GeV$^2$. The mixing parameter 
$\beta$ can be chosen to minimize the overlap with the excited states
and broaden the valid Borel window. Here we adopt $\beta = 0$ for the 
sum rules from the mixed correlator and $\beta = -1.2$ for the sum 
rules from the 1/2 correlator, as determined in Ref.~\cite{leinweber97}.

The vacuum susceptibilities arise from the response of condensates
to the external field. It is easy to show that to first order in
the external field 
\begin{equation}
\langle\overline{q}{\cal O}_{\alpha\beta} q\rangle_{_Z}
= Z^{\gamma\delta}\,  \lim_{p_\lambda\rightarrow 0}
i \int d^4 x e^{i p\cdot y}
\langle T[\sum_{q} g_q\,\overline{q}(y)\sigma_{\gamma\delta} q(y),
{\cal O}_{\alpha\beta}(0)]\rangle\ ,
\label{sus-def-gen}
\end{equation}
where ${\cal O}_{\alpha\beta} = \{\overline{q}\sigma_{\alpha\beta}q,\,
\overline{q}g_s G_{\alpha\beta} q,\, \overline{q} g_s \widetilde{G}
\gamma_5 q \}$. The two-point correlation function in Eq.~(\ref{sus-def-gen}) 
can be decomposed into 
\begin{eqnarray}
i\int d^4 x e^{ip\cdot y}
\langle&& T[\sum_{q} g_q\, \overline{q}(y)\sigma_{\gamma\delta}q(y),
{\cal O}_{\alpha\beta}(0)]\rangle
=\left(g_{\gamma\alpha}\, g_{\delta\beta}-
g_{\gamma\beta}\, g_{\delta\alpha}\right) \Pi^{\cal O}_1(p^2)
\nonumber
\\*[7.2pt]
&&+\left(g_{\gamma\alpha}\, p_\delta p_\beta
-g_{\delta\alpha}\, p_\gamma p_\beta
-g_{\gamma\beta}\, p_\delta p_\alpha 
+g_{\delta\beta}\, p_\gamma p_\alpha\right)
\Pi^{\cal O}_2(p^2)\ .
\label{sus-tpt}
\end{eqnarray}
Neglecting isospin breaking effect and strangeness contribution and
using Eqs.~(\ref{sus-chi}--\ref{sus-xi}), we find
\begin{equation}
\chi a
= -8\pi^2 \Pi^\chi_1(0)\ , 
\hspace*{0.8cm}
\kappa a
= -8\pi^2 \Pi^\kappa_1(0)\ ,
\hspace*{0.8cm}
\xi a
= -8\pi^2 i \Pi^\xi_1(0)\ .
\label{xi-def}
\end{equation} 
These definitions are equivalent to those given
in Ref.~\cite{he96}. It is worthwhile emphasizing that the perturbative
contribution to $\Pi^{\cal O}_1(0)$ has to be subtracted 
as it has already been included through the perturbative 
quark propagator in the presence of the external field (see Appendix). 

Obviously, it is difficult to evaluate $\Pi^{\cal O}_1(0)$ 
directly from QCD at this stage. The QCD sum-rule method has been
invoked to estimate $\Pi^{\cal O}_1(0)$ previously. The results
given in Ref.~\cite{he96} are $\chi a \approx -0.15$GeV$^2$ and
$\kappa a = \xi a \approx 0.1$GeV$^4$. Recently, the authors of 
Ref.~\cite{belyaev97} have argued that the estimate of the vacuum
tensor susceptibility $\chi$ in
Ref.~\cite{he96} is very rough, and their detailed analyses lead 
to a value for $\chi$ which has opposite sign and much larger 
($\sim 4$ times larger) magnitude. Given this uncertainty, we 
will consider the $\chi$ values in the following range
\begin{equation}
-0.5\, \mbox{GeV}^2\, \leq \chi a \leq 1.5\, \mbox{GeV}^2\ .
\label{chi-value}
\end{equation}
Since $\kappa$ and $\xi$ are small and the sum rules are relatively 
insensitive to their values, we use the values of $\kappa$ and $\xi$
given in Ref.~\cite{he96} for simplicity.

The extraction of the nucleon tensor charge from the
sum rules  also requires the knowledge of 
the nucleon mass $M_N$, the couplings $\lambda_1$ and $\lambda_2$, and 
the continuum threshold $s_0$ (for the sum rules in the usual formalism). 
These parameters can be determined from the corresponding mass sum rules. 
Here we take the central values obtained in Ref.~\cite{leinweber97}
\begin{equation}
M_N = 0.96 \mbox{GeV}\ ,\hspace*{0.8cm}
\widetilde{\lambda}_1 \widetilde{\lambda}_2 = 0.41 \mbox{GeV}^6\ , 
\hspace*{0.8cm}
s_0 = 1.69 \mbox{GeV}^2\ ,
\label{m-sr-mix-result}
\end{equation}
for the sum rules from the mixed correlator and 
\begin{equation}
M_N = 1.17 \mbox{GeV}\ ,\hspace*{0.8cm}
\widetilde{\lambda}_1^2  = 0.20 \mbox{GeV}^6\ , 
\hspace*{0.8cm}
s_0 = 2.34 \mbox{GeV}^2\ ,
\label{m-sr-usual-result}
\end{equation}
for the sum rules from the 1/2 correlator. Note that the nucleon mass 
in Eq.~(\ref{m-sr-mix-result}) is very close to its experimental value.
It is, however,  much larger than the experimental value in 
Eq.~(\ref{m-sr-usual-result}). In the literature it is often found 
that the nucleon mass is fixed at its experimental value and the 
mass sum rules are used to determine the couplings and continuum 
threshold. This may introduce artificial errors. Moreover, 
when the same criteria are applied to the analyses of the nucleon 
mass sum rules, we find that the continuum threshold cannot be searched 
successfully when the nucleon mass is held fixed at its experimental
value. 
\subsection{Results}

Let us start from the sum rules in the usual formalism,  
Eqs.(\ref{sr-mix-1}--\ref{sr-mix-2}) and (\ref{sr-usual}).
The nucleon tensor charge $g_{_{\rm T}}$ and the transition
strength $A_i$ are extracted from the sum rules, with the 
continuum threshold fixed at its value for the corresponding 
nucleon mass sum rule. We find that for the $\chi$ values 
interested here there does not exist valid Borel window 
for the sum rule Eq.~(\ref{sr-mix-2}) according to the criteria
set above; it is thus invalid. 

\begin{figure}[t]
\begin{center}
\epsfysize=11.7truecm
\leavevmode
\setbox\rotbox=\vbox{\epsfbox{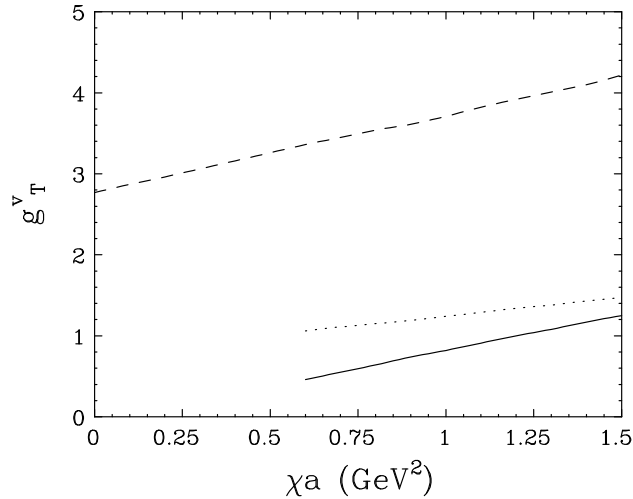}}\rotl\rotbox
\end{center}
\caption{Sum-rule predictions for the isovector
nucleon tensor charge $g^v_{_{\rm T}}$ as functions of $\chi$. The
solid and dashed curves correspond to the sum rules  
Eqs.~(\protect{\ref{sr-mix-1}}) and (\protect{\ref{sr-usual}}), 
respectively. The dotted curve is also from the sum rule 
Eq.~(\protect{\ref{sr-usual}}) but with $M_N = 0.94$GeV, 
$\widetilde{\lambda}_1^2 = 0.26$GeV$^6$, and $s_0 = 2.3$GeV$^2$
as given in Ref. \protect\cite{he96}.%
}
\label{fig-1}
\end{figure}

\begin{figure}[t]
\begin{center}
\epsfysize=11.7truecm
\leavevmode
\setbox\rotbox=\vbox{\epsfbox{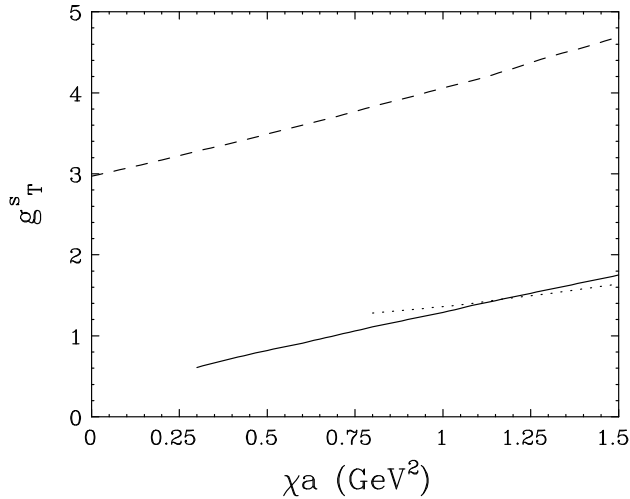}}\rotl\rotbox
\end{center}
\caption{Sum-rule predictions for the isoscalar
nucleon tensor charge $g^s_{_{\rm T}}$ as functions of $\chi$. The
solid and dashed curves correspond to the sum rules 
Eqs.~(\protect{\ref{sr-mix-1}}) 
and (\protect{\ref{sr-usual}}), respectively.
The dotted curve is also from the sum rule 
Eq.~(\protect{\ref{sr-usual}}) but with $M_N = 0.94$GeV, 
$\widetilde{\lambda}_1^2 = 0.26$GeV$^6$, and $s_0 = 2.3$GeV$^2$
as given in Ref. \protect\cite{he96}.%
}
\label{fig-2}
\end{figure}

The sum rule  Eq.~(\ref{sr-mix-1}) works only for a limited range 
of $\chi$ values. The predicted isovector ($g^v_{_{\rm T}}$) and 
isoscalar ($g^s_{_{\rm T}}$) nucleon tensor charges as functions 
of $\chi$ are plotted in Figs. \ref{fig-1} and \ref{fig-2}, 
respectively. Both $g^v_{_{\rm T}}$ and $g^s_{_{\rm T}}$ increase
as $\chi$ increases with essentially the same rate. This indicates
that $\delta u$ strongly depends on $\chi$ while $\delta d$ has
much weaker dependence on $\chi$. One can see that $g^s_{_{\rm T}} 
> g^v_{_{\rm T}} > 0$, implying $\delta u> \delta d > 0$. We also find
that the size of the valid Borel window shrinks with decreasing
$\chi$.

The sum rule  Eq.~(\ref{sr-usual}) is valid for all the $\chi$
values considered here. Its predictions for the nucleon tensor
charges are also given in Figs.~\ref{fig-1} and \ref{fig-2}.
The predicted tensor charges increase as $\chi$ increases
with a similar rate as those obtained from Eq.~(\ref{sr-mix-1}).
The predictions again show $\delta u > \delta d > 0$, but $\delta d$ has 
stronger $\chi$ dependence than that found from Eq.~(\ref{sr-mix-1}).
For a given $\chi$ value, the size of the valid Borel window is 
in general larger than that for Eq.~(\ref{sr-mix-1}).

One notices that the predictions from  Eq.~(\ref{sr-usual}) are significantly 
larger than those from  Eq.~(\ref{sr-mix-1}). This partially attributes to
the large nucleon mass and small coupling [see Eq.~(\ref{m-sr-usual-result})]
used in  Eq.~(\ref{sr-usual}).
To further illustrate this point, we also displayed in Figs.~\ref{fig-1}
and \ref{fig-2} the predictions obtained from the sum rule 
Eq.~(\ref{sr-usual}), 
with $M_N = 0.94$GeV, $\widetilde{\lambda}_1^2 = 0.26$GeV$^6$, 
and $s_0 = 2.3$GeV$^2$ as quoted in Ref.~\cite{he96}. In this case, 
the sum rule  Eq.~(\ref{sr-usual}) only works for a limited range 
of $\chi$ values and the results are much closer to those 
from  Eq.~(\ref{sr-mix-1}).

We now turn to the sum rules in the alternative formalism, 
Eqs.~(\ref{sr-mix-1a}--\ref{sr-mix-2a}) and (\ref{sr-usual-a}).
As emphasized earlier, the continuum threshold $s_0$ must be 
treated as an unknown parameter to be extracted from the sum rules
along with the tensor charges. We find that the sum rule  Eq.~(\ref{sr-mix-2a})
fails to have a valid Borel window and is thus invalid. The sum 
rules  Eqs.~(\ref{sr-mix-1a}) and (\ref{sr-usual-a}) are also invalid
for the case of isoscalar tensor charge. For isovector tensor charge,
the sum rule  Eq.~(\ref{sr-mix-1}) only works for small $\chi$ 
values ($\chi \leq 0.1$) and its prediction is negative
($\sim -1.02$--- $-0.16$). These results are much smaller than the
corresponding ones in the usual formalism. On the other hand, the sum rule
Eq.~(\ref{sr-usual-a}) is only valid for large $\chi$ value ($\chi
\geq 0.8$) and the predictions are slightly larger than those
found from  Eq.~(\ref{sr-usual}).  
\subsection{Discussion}

\begin{figure}[t]
\begin{center}
\epsfysize=11.7truecm
\leavevmode
\setbox\rotbox=\vbox{\epsfbox{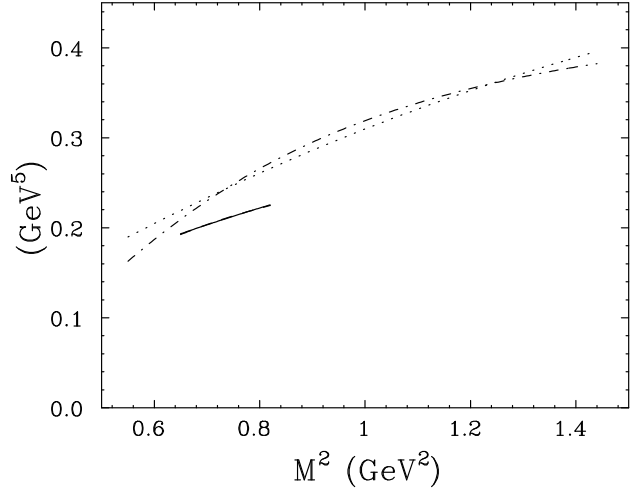}}\rotl\rotbox
\end{center}
\caption{The left- and right-hand sides of the sum rules
Eqs.~(\protect{\ref{sr-mix-1}}) and (\protect{\ref{sr-usual}}) as 
functions of the Borel
mass for $\chi = 1.0$, with $g^v_{_{\rm T}}$ and $A_i$ extracted 
from the sum rules. The four curves correspond to the left-
(solid curve) and right-hand (dashed curve) 
sides of Eq.~(\protect{\ref{sr-mix-1}})
and the left- (dotted curve) and right-hand (dot-dashed curve) sides of
Eq.~(\protect{\ref{sr-usual}}), respectively}
\label{fig-3}
\end{figure}

\begin{figure}[b]
\begin{center}
\epsfysize=11.7truecm
\leavevmode
\setbox\rotbox=\vbox{\epsfbox{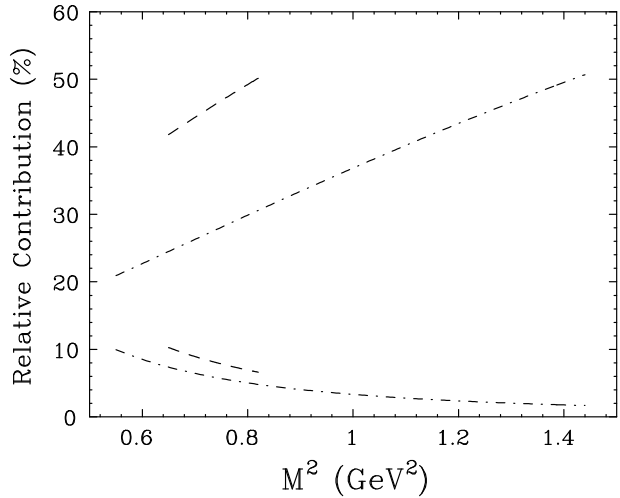}}\rotl\rotbox
\end{center}
\caption{Relative contributions of the transition plus continuum
and the highest order OPE terms to the sum rules as functions of 
the Borel mass. The dashed and dot-dashed curves correspond to
the sum rules Eqs.~(\protect{\ref{sr-mix-1}}) and 
(\protect{\ref{sr-usual}}), respectively.}
\label{fig-4}
\end{figure}

To see how well the valid sum rules work, we have plotted in Fig.~\ref{fig-3}
the left- and right-hand sides of the sum rules Eqs.~(\ref{sr-mix-1})
and (\ref{sr-usual}) for $\chi = 1.0$, with $g^v_{_{\rm T}}$ and
$A_i$ extracted from the sum rules. The corresponding valid Borel windows 
and the relative contributions of the transition plus continuum and the 
highest order term in the OPE are displayed in Fig.~\ref{fig-4}. One notices
that the sum rule Eq.~(\ref{sr-usual}) is valid in a broad Borel regime
where the transition plus continuum contributes about $20\%$ at the 
lower bound and the required $50\%$ at the upper bound. The sum rule 
Eq.~(\ref{sr-mix-1}) has a much smaller valid Borel regime where the 
contributions from the transition and continuum are greater than 
$40\%$ throughout the valid Borel window. On the other hand,  the overlap 
of the two sides of Eq.~(\ref{sr-mix-1}) is better than that of 
Eq.~(\ref{sr-usual}). This pattern is seen at other $\chi$ values. 
It is therefore hard to compare the reliabilities of these two sum 
rules directly. 

As shown in Eq.~(\ref{old-form}), the contribution from the transitions
is in general a complicated function of the Borel mass, which is 
crudely modeled by a constant ($A_i$) in the usual formalism. This 
will change the curvature of the phenomenological side as a function
of the Borel mass and may lead to errors to the extracted nucleon
tensor charges. It is unclear how to reliably estimate the size of 
these errors within the criteria discussed in the present paper, 
though some estimates have been given in the study of other nucleon 
properties \cite{jin97,ioffe95}.

Moreover, the spectral parameters (mass, coupling, and continuum
threshold) appearing in the nucleon mass sum rules also enter the 
sum rules for the nucleon tensor charge. This means that the 
uncertainties associated with these parameters will give rise to
additional uncertainties, in addition to the uncertainties in the
sum rules for the tensor charge themselves. This is a general drawback of
the external field sum-rule approach, and we have demonstrated this point
explicitly here. Clearly, one needs accurate knowledge of the
vacuum spectral parameters in order to extract the tensor charges
cleanly, which is difficult to achieve given the current implementation
of the QCD sum-rule approach. Therefore, we expect large uncertainties 
associated with the predictions of the valid sum rules in the usual 
formalism presented above. 
      
In the alternative formalism, the contribution of the transitions
is absorbed into the continuum. It has been argued recently 
by one of us \cite{jin97} that this alternative formalism may
have the potential to improve the reliability of external-field 
sum-rule calculations, as the transitions are explicitly modeled 
by the modified continuum model. However, we have found here that 
the sum rules in the alternative formalism fail in most cases 
interested here. This may indicate that the information contained
in the truncated OPE is not enough and/or the transition and continuum 
give rise to too much background noise to isolate the information 
about the nucleon tensor charge.  This might also suggest that the 
reason for the sum rules in the usual formalism to work is because 
the transition contribution is approximated by a constant. 

We have emphasized throughout this paper that in the alternative formalism
the continuum threshold has to be treated as a unknown parameter because the
continuum model is modified relative to that in the usual formalism
and contains the transition contribution. Fixing the continuum threshold 
at its value for the corresponding mass sum rule may introduce arbitrary 
artificial effect to the extracted tensor charges. In Ref.~\cite{he96},
the sum rule Eq.~(\ref{sr-usual-a}) is used in extracting the nucleon
tensor charges but with the continuum threshold fixed at its value
for the nucleon mass sum rule. However, we have found that this sum
rule is invalid for $\chi a =-0.15$GeV$^2$ as used in Ref.~\cite{he96},
and hence the results obtained from this sum rule are unreliable.

\begin{figure}[t]
\begin{center}
\epsfysize=11.7truecm
\leavevmode
\setbox\rotbox=\vbox{\epsfbox{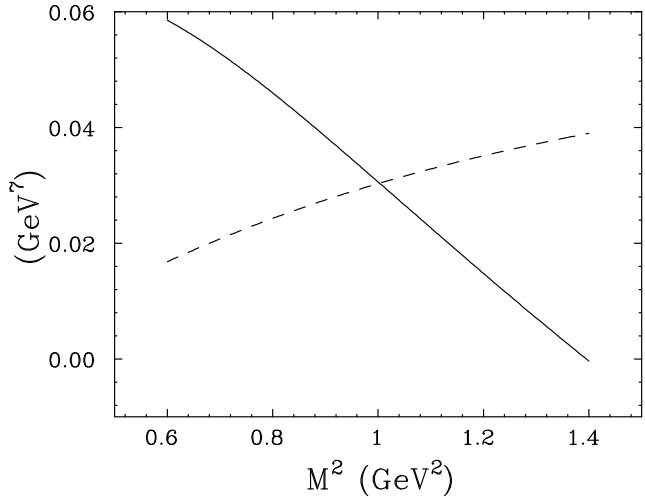}}\rotl\rotbox
\end{center}
\caption{The left- (solid curve) and right-hand (dashed curve) sides 
of the sum rule (\protect{\ref{sr-usual-a}}) as functions of the Borel 
mass, with the use of $M_N = 0.94$GeV, 
$\widetilde{\lambda}_1^2 = 0.26$GeV$^6$, $\chi a =-0.15$GeV$^2$,
and the continuum threshold fixed at $s_0 = 2.3$GeV$^2$ 
as quoted in Ref. \protect\cite{he96}. The sum rule is not optimized 
and $g^s_{_{\rm T}} = 0.3$ is adopted.}
\label{fig-5}
\end{figure}

In their analyses, the authors of Ref.~\cite{he96} express the tensor 
charge as a function of the Borel mass by dividing both sides of 
Eq.~(\ref{sr-usual-a}) by $M_N \widetilde{\lambda}_1^2
e^{-M_N^2/M^2}$ and find the tensor charge changes quickly with
the Borel mass. This is actually an indication of the poor overlap of the
two sides of the sum rule. To illustrate this point, we have plotted 
in Fig.~\ref{fig-5} the left- and right-hand sides of 
Eq.~(\ref{sr-usual-a}), with the use of  $M_N = 0.94$GeV, 
$\widetilde{\lambda}_1^2 = 0.26$GeV$^6$, $\chi a =-0.15$GeV$^2$,
and the continuum threshold fixed at $s_0=2.3$GeV$^2$ as quoted 
in Ref.~\cite{he96}. Here following Ref.~\cite{he96}, the sum rule
is not optimized and the prediction for the nucleon tensor charge 
$g^s_{_{\rm T}} \simeq 0.3$ is obtained by simply taking $M^2 = 1$GeV$^2$.
We see that the two sides of the sum rule does not match at all
and they only coincide at the point $M^2 = 1$GeV$^2$, which is picked 
by hand. So, the prediction can be arbitrary as one may choose an arbitrary
value for the Borel mass. 

\section{Summary and conclusion}
\label{summary}

In this paper, we have derived two new QCD sum rules for the
nucleon tensor charge from the mixed correlator of spin-1/2 
and spin-3/2 interpolating fields. Both the usual formalism
and the alternative formalism were adopted to handle the 
transitions between the ground state nucleon and 
excited states. We analyzed these new sum rules, along with the
chiral-odd sum rule obtained from the usual 1/2 correlator, with respect
to specific criteria for monitoring the contaminations arising from
the transitions and continuum and the convergence of the OPE. 
An emphasis was put on revealing the validity and reliability 
of various sum rules.

We found that in the usual formalism one of the two new sum rules
is invalid and the other works only for certain values of $\chi$.
The predictions of the latter for moderate $\chi$ values
(0.5 GeV$^2 < \chi a <$ 1.0 GeV$^2$) are 
\begin{equation}
\delta u \sim 0.67 - 1.06\ , 
\hspace*{1cm}
\delta d \sim 0.23 - 0.24\ .
\label{res-sr-mix-1}
\end{equation}
This shows that $\delta u$ has strong dependence on the 
vacuum tensor susceptibility $\chi$ and $\delta d$ is
insensitive to $\chi$. The sum rule from
the 1/2 correlator is valid for all the $\chi$ values
considered, and the predictions for the moderate
$\chi$ are
\begin{equation}
\delta u \sim 3.48 - 3.89\ , 
\hspace*{1cm}
\delta d \sim 0.12 - 0.18\ .
\label{res-sr-usual}
\end{equation}
It should be warned that the uncertainties quoted here
only reflect the uncertainties from the $\chi$ value. 
There are large uncertainties associated with the usual 
formalism arising from the treatment of the 
transitions, the spectral parameters in the mass sum rules, and 
the vacuum condensates and other susceptibilities. Without 
evaluating these uncertainties, we cannot make
a critical comparison with the results from other models
and approaches as these uncertainties may change these 
results significantly. Therefore, the above results should 
be treated only as benchmark (at most). Nevertheless,
our results appear to support $\delta u >
\delta d > 0$. This behavior is different from that
found in Rebs.~\cite{kim96,schmidt97,aoki96} and the
behavior of the quark spin structure of the nucleon
($\Delta u > 0$, $\Delta d < 0$).

The sum rules in the alternative formalism
are invalid in most cases of interest. In particular, for isoscalar tensor
charge all three sum rules are invalid. For isovector tensor
charge, one of the two new sum rules is invalid for all the 
$\chi$ values, and the other only works marginally for small 
$\chi$ values and its predictions for the tensor charge are
much smaller than those given in Eq.~(\ref{res-sr-mix-1}). 
The sum rule from the 1/2 correlator works only for large 
$\chi$ values and its predictions are similar to those listed in 
Eq.~(\ref{res-sr-usual}). It is also found that the sum rule
Eq.~(\ref{sr-usual-a}) with $\chi a =-0.15$GeV$^2$, studied
in the previous works, is invalid. 

In conclusion, the QCD sum rules for the nucleon tensor charge
work only in limited cases and their predictions are expected
to have large uncertainties. It is important to examine the
validity and reliability of the sum rules before the extraction
can be made. Without such an examination, the extracted
results are likely to be arbitrary and misleading. To improve
the performances of the sum rules for the nucleon tensor charge,
one should have more accurate evaluation of the vacuum
susceptibilities and vacuum condensates, more precise
knowledge of the mass sum-rule spectral parameters, 
higher order terms in the OPE, and better approaches for 
treating the transitions.

\acknowledgements

We thank Bob Jaffe for helpful discussions, and B. L. Ioffe, 
Kai Henchen and Hanxin He for useful correspondences.
This work is supported in part by funds provided by the U.S.
Department of Energy (D.O.E.) under cooperative 
research agreement \#DF-FC02-94ER40818.    

\appendix
\label{app}
\section*{}

In this appendix, we give some ingredients for calculating 
the QCD side as obtained from carrying out the OPE. Further
details of the technique can be found in the literature.
As usual, we work to the leading order in perturbative theory and 
neglect the up and down current quark masses, the isospin
breaking effect, and the strangeness contribution. 
It is convenient to calculate the Wilson 
coefficients of the OPE in coordinate space and then perform
the Fourier transformation to momentum space. To first order 
in the external field, the quark propagator without background
gluon field can be expressed as
\begin{eqnarray}
\langle && T[q^a_i(x) q^b_j(0)]\rangle_{_Z}
={i\over 2\pi^2}\,{\hat{x}_{ij}\over x^4}\,\delta^{ab}
-{\langle\overline{q}q\rangle\over 12}\,\delta_{ij}\,\delta^{ab}
+{x^2\over 192}\,\langle\overline{q} g_s\sigma\cdot G q\rangle 
\,\delta_{ij}\,\delta^{ab}
\nonumber
\\*[7.2pt]
&&-{x^4\over 27648}\,\langle\overline{q}q\rangle
\,\langle g_s^2 G^2\rangle\, \delta_{ij}\,\delta^{ab}
-{g_q\over 4\pi^2}\,Z^{\alpha\beta}\, {[\hat{x}\sigma_{\alpha\beta}
\hat{x}]_{ij}
\over x^4}\,\delta^{ab}
\nonumber
\\*[7.2pt]
&& - {g_q\over 24}\, \chi\,\langle\overline{q} q\rangle 
\,Z^{\alpha\beta}\, (\sigma_{\alpha\beta})_{ij}\,\delta^{ab}
-i\, {g_q\over 48}\, \langle\overline{q}q\rangle\, Z^{\alpha\beta}
\,(\hat{x}\sigma_{\alpha\beta}+\sigma_{\alpha\beta}\hat{x})_{ij}
\,\delta^{ab}
\nonumber
\\*[7.2pt]
&&+{g_q\over 288}\,\langle\overline{q}q\rangle
\,Z^{\alpha\beta}\,\left[
x^2\, (\kappa + 2\xi)\, \sigma_{\alpha\beta}
+2\, x_\beta\, x^\gamma\, (\kappa-\xi)\,\sigma_{\alpha\gamma}
\right]_{ij}\,\delta^{ab} +\cdots\ ,
\label{prop}
\end{eqnarray}
where $\{i,j\}$ and $\{a,b\}$ are Dirac and color indices, respectively,
and the fifth term is the perturbative propagator in the presence of the
external field. Note that the sign in front of $\xi$ differs from that 
given in Ref.~\cite{he96}. Here we adopted the conventions of Itzykson
and Zuber for $\epsilon_{\alpha\beta\rho\lambda}$ while Ref.~\cite{he96}
has not given the conventions explicitly.

In the fixed-point gauge, the quark propagator in the presence of background
gluon field is given by
\begin{eqnarray}
\langle && T[ q^a_i(x) \overline{q}^b_j(0)]\rangle_{_{\rm Z}} =
{i\over 32\pi^2}\, g_s G^n_{\rho\lambda}{\,[\hat{x}\sigma^{\rho\lambda}
+\sigma^{\rho\lambda}\hat{x}]_{ij}\over x^2} 
\,\left({\lambda^n\over 2}\right)^{ab}
\nonumber
\\*[7.2pt]
&&
+{1\over 288}\, \langle\overline{q}q\rangle
\,g_s G^n_{\rho\lambda}\, \left[x^2\sigma^{\rho\lambda} - 2 x_\delta x^\lambda
\sigma^{\delta\rho}\right]_{ij}\, \left({\lambda^n\over 2}\right)^{ab}
\nonumber
\\*[7.2pt]
&&
+{g_q\over 64\pi^2}\,  g_s G^n_{\rho\lambda}\, Z_{\alpha\beta}
\,\Biggl\{ -{\sigma^{\rho\lambda}\hat{x}\sigma^{\alpha\beta}\hat{x}
+\hat{x}\sigma^{\alpha\beta}\hat{x}\sigma^{\rho\lambda}\over x^2}
\nonumber
\\*[7.2pt]
&&
+i\,\left(\gamma^\rho\sigma^{\alpha\beta}\gamma^\lambda
-\gamma^\lambda\sigma^{\alpha\beta}\gamma^\rho\right)
\,\left[\ln\left(-{x^2\Lambda^2\over 4}\right)+
2\gamma_{_{\rm EM}}\right]\Biggr\} _{ij}
\,\left({\lambda^n\over 2}\right)^{ab}
\nonumber
\\*[7.2pt]
&&
-{g_q\over 6912}\, \langle g_s^2 G^2\rangle 
\,Z_{\alpha\beta}
\,\left[ \ln\left(-{x^2 \Lambda^2\over 4}\right)+2\gamma_{_{\rm EM}}\right]
\left[\hat{x}\sigma^{\alpha\beta}\hat{x}\right]_{ij}\, \delta^{ab}
\nonumber
\\*[7.2pt]
&&
+i{g_q\over 288}\,\langle\overline{q}q\rangle
\, g_s G^n_{\rho\lambda}
\,\left[Z^{\rho\lambda}\, x^2\hat{x}
+2 Z^{\delta\rho}\, x^2 x^\lambda \,\gamma_\delta
+2 i\, \widetilde{Z}^{\delta\rho}\, x^2 x_\delta\, \gamma_5\gamma^\lambda
\right]_{ij}\, \left({\lambda^n\over 2}\right)^{ab}
\nonumber
\\*[7.2pt]
&&
+i{g_q\over 24}\, \chi\, \langle\overline{q}q\rangle
 \,g_s G^n_{\rho\lambda}
\,Z_{\alpha\beta}\, x^\beta x^\lambda \,[\sigma^{\alpha\rho}]_{ij}
\,\left({\lambda^n\over 2}\right)^{ab}
\nonumber
\\*[7.2pt]
&&
+{g_q\over 144}\, \chi\, \langle\overline{q}q\rangle
\, g_s G^n_{\rho\lambda}
\,\left( Z^{\rho\lambda}\, x^2 -Z^{\rho\delta}\, x_\delta x^\lambda\right)
\delta_{ij}
\,\left({\lambda^n\over 2}\right)^{ab}
+\cdots\ ,
\label{prop-1}
\end{eqnarray}
where the third and fourth terms arise from the presence of both
the external and background gluon fields. The last three
terms have been ignored in Ref.~\cite{he96}, which gives rise to
the differences in the last four terms in the sum rule 
Eq.~(\ref{sr-usual-a}). In the calculations of the terms
proportional to $ab$, we have neglected the fourth and fifth
terms of Eq.~(\ref{prop-1}). We have found that the inclusion 
of such terms only make tiny refinement to the sum rules.

We also need the following expression for the expectation value
of the quark and gluon field product in the presence of the external
field \cite{ioffe84}
\begin{eqnarray}
\langle  q^a_i \overline{q}^b_j g_s G^n_{\alpha\beta}\rangle_{_{\rm Z}}
=&& - {1\over 192}\langle\overline{q} g_s \sigma\cdot G q\rangle
[\sigma_{\alpha\beta}]_{ij} \left({\lambda^n\over 2}\right)^{ab}
\nonumber \\
&&
-{1\over 16}\left[
\kappa Z_{\alpha\beta} + {1\over 2} i \xi
\epsilon_{\alpha\beta\rho\lambda} Z^{\rho\lambda} \gamma_5
\right]_{ij} \left({\lambda^n\over 2}\right)^{ab}\ .
\label{prop-extra}
\end{eqnarray}

Now one can easily carry out the calculations of the correlators 
by expressing the correlators as \cite{leinweber97}
\begin{eqnarray}
\langle  T[&&\eta^{1/2}_\mu(x)  \overline{\eta}^{3/2}_\nu(0)]\rangle_{_Z}
=
\nonumber \\ & &
\epsilon^{abc}\epsilon^{a^\prime b^\prime c^\prime}\Biggl\{\;
\beta \gamma_\mu S^{aa^\prime}_u \gamma_\nu \sigma_{\rho\lambda}
\mbox{Tr}\left[ S^{bb^\prime}_d \sigma^{\rho\lambda} 
C {S^{cc^\prime}_u}^T C\right]
+\gamma_\mu\gamma_5  S^{aa^\prime}_u \gamma_\nu \sigma_{\rho\lambda}
\mbox{Tr}\left[\gamma_5 S^{bb^\prime}_d \sigma^{\rho\lambda}
C {S^{cc^\prime}_u}^T C\right]
\nonumber \\ & &
-2\gamma_\mu\gamma_5  S^{aa^\prime}_u \sigma_{\rho\lambda}
C {S^{cc^\prime}_u}^T C \gamma_5 S^{bb^\prime}_d
\gamma_\nu \sigma^{\rho\lambda}
-\gamma_\mu\gamma_5  S^{aa^\prime}_u \sigma_{\rho\lambda}
C {S^{cc^\prime}_d}^T C \gamma_5 S^{bb^\prime}_u
\gamma_\nu \sigma^{\rho\lambda}
\nonumber \\ & &
-2\beta\gamma_\mu  S^{aa^\prime}_u \sigma_{\rho\lambda}
C {S^{cc^\prime}_u}^T C  S^{bb^\prime}_d
\gamma_\nu \sigma^{\rho\lambda}
-\beta\gamma_\mu  S^{aa^\prime}_u \sigma_{\rho\lambda}
C {S^{cc^\prime}_d}^T C S^{bb^\prime}_u
\gamma_\nu \sigma^{\rho\lambda}\; \Biggr\},
\label{master-mix}
\\*[7.2pt]
%
\langle T[&&\eta^{1/2}(x) \overline{\eta}^{1/2}(0)]\rangle_{_Z}
=
\nonumber \\&&
-\epsilon^{abc}\epsilon^{a^\prime b^\prime c^\prime} \Biggl\{\;
\beta^2 \gamma_5 S^{aa^\prime}_u \gamma_5
\mbox{Tr}\left[ C {S^{cc^\prime}_d}^T C S^{bb^\prime}_u\right]
+S^{aa^\prime}_u \mbox{Tr}\left[C {S^{cc^\prime}_d}^T C 
\gamma_5 S^{bb^\prime}_u \gamma_5\right]
\nonumber \\ & &
+\beta \gamma_5  S^{aa^\prime}_u 
\mbox{Tr}\left[ C {S^{cc^\prime}_u}^T C S^{bb^\prime}_d \gamma_5\right]
+\beta S^{aa^\prime}_u \gamma_5
\mbox{Tr}\left[ C {S^{cc^\prime}_u}^T C \gamma_5 S^{bb^\prime}_d\right]
\nonumber \\ & &
+ S^{aa^\prime}_u \gamma_5
C {S^{cc^\prime}_d}^T C  \gamma_5 S^{bb^\prime}_u
+\beta\gamma_5  S^{aa^\prime}_u \gamma_5
C {S^{cc^\prime}_d}^T C S^{bb^\prime}_u
\nonumber \\ & &
+\beta  S^{aa^\prime}_u
C {S^{cc^\prime}_d}^T C \gamma_5 S^{bb^\prime}_u \gamma_5
+\beta^2 \gamma_5  S^{aa^\prime}_u 
C {S^{cc^\prime}_d}^T C S^{bb^\prime}_u \gamma_5\;\Biggr\},
\label{master-1/2}
\end{eqnarray}
with the quark propagator $S^{ab}_q(x,0;Z) \equiv \langle 
T[q^a(x)\overline{q}^b(0)]\rangle_{_{\rm Z}}$ given above. Since only
linear response of the correlators is concerned, one only 
needs to keep the terms linear in the external field.

Finally, we observe that the Borel transformation of $\ln(-q^2)/q^2$
has not been carried out correctly in Ref.~\cite{he96}. So, we give 
the correct formula \cite{wilson87}
\begin{equation}
{\cal B}\left[
{\ln(-q^2)\over q^2}
\right]
= -\ln(M^2) + \gamma_{_{\rm EM}}\ ,
\label{borel-eg}
\end{equation}
from which we get
\begin{equation}
{\cal B}\left[
\left(M_N^2 - q^2\right) {\ln\left(\Lambda^2/-q^2\right)\over q^2}
\right]
= M_N^2 \left[\ln\left({M^2\over \Lambda^2}\right) -\gamma_{_{\rm EM}}\right]
-M^2\  .
\label{borel-ega}
\end{equation}
It is then easy to show that the second term of Eq. (20) in Ref.~\cite{he96}
is not the correct Borel transform of the second term of Eq. (17).


\end{document}